\documentclass[twocolumn,aps,floatfix]{revtex4}
\usepackage{graphicx}
\usepackage{amsmath}
\usepackage{float}
\usepackage{color}
\begin{document}

\title{Stratified construction of neural network based interatomic models for multicomponent materials}

\author{Samad Hajinazar, Junping Shao, Aleksey N. Kolmogorov}

\affiliation{Department of Physics, Applied Physics and Astronomy, Binghamton University, State University of New York,
PO Box 6000, Binghamton, New York 13902-6000, USA}

\date{\today}

\begin{abstract}
{Recent application of neural networks (NNs) to modeling interatomic
  interactions has shown the learning machines' encouragingly accurate
  performance for select elemental and multicomponent systems. In this
  study, we explore the possibility of building a library of NN-based
  models by introducing a hierarchical NN training. In such a
  stratified procedure NNs for multicomponent systems are obtained by
  sequential training from the bottom up: first unaries, then
  binaries, and so on. Advantages of constructing NN sets with shared
  parameters include acceleration of the training process and intact
  description of the constituent systems. We use an automated
  generation of diverse structure sets for NN training on density
  functional theory-level reference energies. In the test case of Cu,
  Pd, Ag, Cu-Pd, Cu-Ag, Pd-Ag, and Cu-Pd-Ag systems, NNs trained in
  the traditional and stratified fashions are found to have
  essentially identical accuracy for defect energies, phonon
  dispersions, formation energies, etc. The models' robustness is
  further illustrated via unconstrained evolutionary structure
  searches in which the NN is used for the local optimization of
  crystal unit cells.}
\end{abstract}

\maketitle


\section{Introduction}

Materials modeling and design require a toolset of diverse methods
capable of describing relevant bonding, electronic, magnetic, optical,
and other properties. In particular, reliable determination of
materials' thermodynamic stability in structure searches or
examination of materials' behavior in molecular dynamics simulations
depend on accurate and efficient evaluation of the total energy and
atomic forces. Given ionic positions, these basic quantities can be
found at different levels of approximation with only partial or no
explicit treatment of the electronic degrees of freedom. Striking a
good balance between computational cost and approximation accuracy is
one of the central goals in methodology development for materials
research  \cite{QUA:QUA24927,QUA:QUA24890,ZTadmor:2011fk,doi:10.1146/annurev-matsci-071312-121610,Tadmor2013298,Tadmor2011,0965-0393-23-7-074008,bartlett2007applications,Jain:2016rt,RevModPhys.87.897}.

Density functional theory (DFT), based on parameterized functionals of
the electron charge density, represents a reliable practical solution
for establishing crystal structure stability for ordered phases with
up to a few hundred atoms. Extensive benchmark studies have shown that
$T=0$ K total energy calculations at the (semi)local DFT approximation
level correctly predict inorganic compound stability in 97\% of
metal-metal \cite{Curtarolo2005163} and 83\% of more complex
metal-boron binaries \cite{VanDerGeest2014184}. Due to the costly $N^3$
scaling of DFT with system size, global ground state structure
searches for systems exceeding several dozen atoms or simulations of
phenomena extending over a few nanometers in size and/or several femtoseconds in time
are prohibitively expensive.

The traditional construction of system-- or interaction-- specific
classical models relies on the physical/chemical understanding of the
bonding mechanisms. Families of bond-order potentials (BOPs),
embedded-atom models (EAM), polarizable ion potentials, and
Lennard-Jones potentials have been developed to treat systems with
dominant covalent, metallic, ionic, and van der Waals types of
interaction, respectively
\cite{PhysRevLett.56.632,PhysRevLett.50.1285,0965-0393-14-5-002,0953-8984-5-17-004,Jones463,Perea2016gf}. The
main advantage of this approach is a relatively small number of
semi-empirical fitting parameters but the rigid functional forms may
in some cases fail to capture the full spectrum of many-body effects
\cite{classbad1,ak06,classbad2}.

An alternative strategy for describing materials properties involves
interpolation of large databases of DFT energies, forces, density of
states (DOS),
etc. \cite{angle,PhysRevB.87.184115,QUA:QUA21398,0953-8984-20-28-285219,DFTNN,Griffin201311920,PhysRevB.89.205118,C3TA13235H,PhysRevE.89.053316,Pilania:2013fk,xref1,xref2,Khorshidi2016310}.
Artificial neural networks (NNs) have proven to be effective machine
learning tools for dealing with multidimensional classification,
control, and interpolation problems in various fields. NNs were first
used to map interatomic potential energy surfaces over two decades
ago. As reviewed by Witkoskie and Doren  \cite{doi:10.1021/ct049976i},
the early studies included parametrization of specific terms in known
potentials or selected geometries. In
Ref. \onlinecite{ak00} radial basis function-based NNs
trained on energies and optionally forces showed good performance
describing graphitic configurations with constant numbers of neighbors
once the atomic environments were broken up into smaller collections
of first and second nearest neighbors. Since NNs require a fixed
number of input components, applicability of these early models was
restricted to specific geometries or covalent framework types.

Major advances made in recent years have helped overcome this
limitation. Behler and Parrinello \cite{PhysRevLett.98.146401} and
Bartok {\it et al.}  \cite{PhysRevB.87.184115} have introduced
so-called descriptors that enable fixed-size interaction-independent
representations of arbitrary atomic environments. Use of these
descriptors to preprocess atomic configurations into inputs for NNs
was found to be effective in select Na, Cu, Si, W, H$_2$O, Cu-ZnO, and
other
systems \cite{PhysRevB.81.184107,PhysRevLett.98.146401,PhysRevB.90.104108,C4CP04751F,PSSB:PSSB201248370}.
Challenges and advances in constructing descriptors are detailed in
several excellent
reviews \cite{0953-8984-26-18-183001,QUA:QUA24890,PhysRevB.87.184115}.

\begin{figure}[t]
\begin{center}
\vspace{0.2cm} \includegraphics[width=72mm]{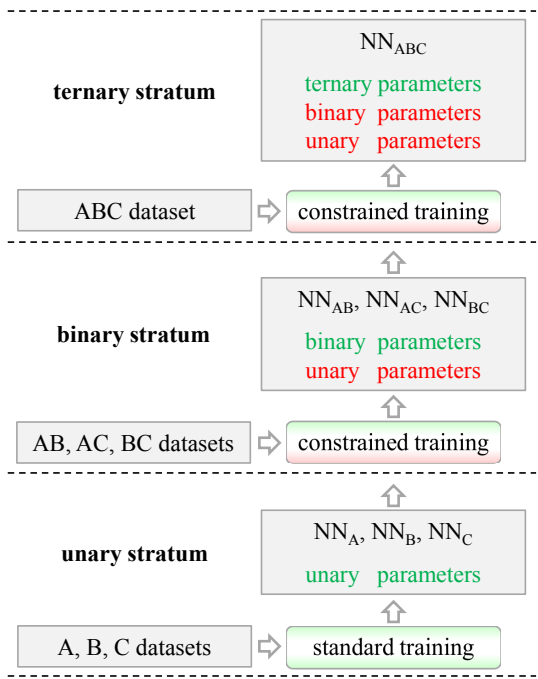}
\caption{(Color online) A flowchart demonstrating what data sets are
  used as well as what parameters are adjusted (in green) and fixed
  (in red) in the stratified training for a ternary system with
  elements A, B, and C. The ternary NN at the top inherits parameters
  from the constituent unary and binary systems. More details are
  given in Sec. V. }
\vspace{-0.5cm}
\end{center}
\end{figure}

In this study we focus on two separate aspects of producing practical
NN-based models for multicomponent systems. First, we employ an
automated structure generation scheme based on an evolutionary
selection procedure that samples parts of the configurational space
typically accessed in global structure searches. Second, we introduce
and test a stratified training scheme as a basis for constructing
extended NN libraries for multicomponent systems. Figure 1 shows a
general flowchart for the training process. Such a hierarchical
approach has been used previously for building classical and
tight-binding models
\cite{PhysRevB.80.165122,PhysRevB.84.155119,jp1048088} but its
potential for constructing NN sets has not been explored. The rest of
this paper is organized as follows: Section II details the employed
data generation protocol; Sec. III reviews the descriptors used in
this study; Sec. IV gives the NN setup basics; Sec. V introduces
the stratified training scheme; Sec. VI illustrates the performance
of the NNs for Cu, Pd, Ag, Cu-Pd, Cu-Ag, Pd-Ag, and Cu-Pd-Ag systems;
finally, Sec. VII summarizes our findings and outlooks.

\section{Generation of training datasets}
\begin{figure*}[t!]
\begin{center}
\includegraphics[width=175mm,angle=0]{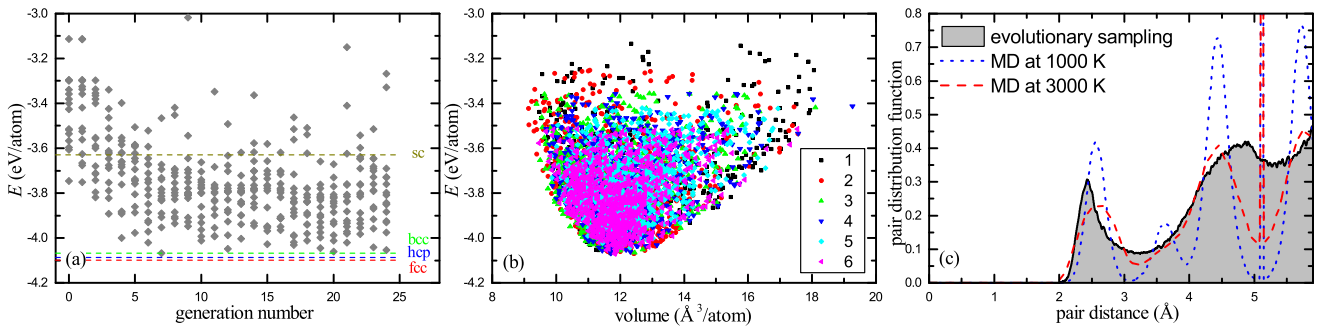}
\caption{Automatically generated structure set for
  Cu. (a) A typical distribution of energies for a 16-structure
  population over 25 generations. Dashed lines show energies for
  select high-symmetry structures. (b) Total distribution of energies
  versus volume for unit cells with 1 to 6 atoms. (c) Pair
  distribution function for 1-6 atom structures (shaded gray) and
  for $3\times 10^4$ 8-atom structures in $2\times 2\times 2$ fcc
  supercells in MD runs at different temperatures (non-shaded lines).}
\end{center}
\end{figure*}

Accuracy and diversity of the data created for NN training are among
the defining factors in the NN overall performance. It is important to
note that developed models inherit the {\it systematic} errors of the
method chosen to calculate structures' total energies, forces, and
stresses. Chemically accurate "post-Hartree-Fock" methods have been
used in some studies \cite{MPN,1367-2630-15-9-095003} but DFT
approximations have been a practical choice for producing reference
values for large data sets. The well-known DFT limitations to
accurately describe the interactions in strongly correlated or van der
Waals systems can be mediated with the DFT+{\it U} \cite{PhysRevB.57.1505,
  PhysRevB.44.943} and the vdW-DF-based
\cite{PhysRevLett.91.126402,vdWinDFT} approaches with little
additional computational cost \cite{
  Burke,Hautier2012,PhysRevB.83.224103}. In some seemingly ordinary
cases, such as the Cu-Au binary, obtaining proper values of formation
energies requires considerably more expensive hybrid functional-level
calculations \cite{PhysRevLett.112.075502}.

For the purpose of investigating the performance of NNs as interatomic
models, all training and testing have been performed on data evaluated
at the same level of DFT approximations. We used the
Perdew-Burke-Ernzerhof (PBE) exchange-correlation functional
 \cite{PBE} within the generalized gradient approximation (GGA)
 \cite{GGA} and projector augmented wave potentials  \cite{PAW}
available in {\small VASP}  \cite{VASP1,VASP2}. The 500 eV energy
cutoff and dense $k$-meshes, $N_{k1} \times N_{k2} \times N_{k3}
\times N_{atoms} \ge 4,000$  \cite{MONKHORST_PACK}, ensured numerical
convergence of the formation energy differences to typically within
1-2 meV/atom.

Generation of diverse structures sampling relevant regions of the
configuration space is critical for training of NNs because of their
notoriously poor function as extrapolators \cite{extrabad}. Previously
proposed strategies include running molecular dynamics (MD), creating
regular meshes of crystal structure parameters, selecting specific
desired geometries, etc. \cite{sample1,PhysRevB.85.045439,sample2}. The
MD-based approach has the advantage of sampling commonly accessible
physical states. However, systems simulated with standard MD tend to
remain in low free-energy basins and configurations along MD
trajectories tend to correlate. A number of algorithms have been
developed that force systems to explore larger regions of the phase
space and that enable selection of more dissimilar structures
 \cite{Tribello03042012,jp0757053,QUA:QUA21398,PSSB:PSSB201248370,PhysRevB.85.045439}.
In particular, an iterative MD-based training protocol proposed in
Ref.  \cite{PhysRevB.85.045439} diversifies training sets by
identifying and adding structures for which pre-trained NNs differ the
most.

Figure 2(c) illustrates a shortcoming in the MD-based generation of
reference data involving small structures with fixed lattice
constants. Heating up small supercells of low-energy structures is an
attractive and cost-efficient way of sampling relevant potential
energy funnels. However, examination of the pair distribution function
reveals an unwanted bias related to the appearance of the same nearest
neighbor distances in all such structures. The area under the sharp
peak at 5.2 \AA\ in the $2\times 2\times 2$ 8-atom fcc supercell makes
up 15\% of the total pair distribution function integrated up to 6
\AA. It is desirable to circumvent the occurrence of such biases
without expanding the structure size. Variable-cell MD is one possible
solution but it is also desirable to avoid long DFT runs even if
performed at reduced accuracy settings.

We used the following guidelines in designing an alternative protocol:
(i) avoid manual selection of structures; (ii) favor low-energy
structures; (iii) ensure diversity; and (iv) find representative
samples commonly seen in global structure searches. These criteria are
satisfied well in mock unconstrained optimization runs driven by an
evolutionary algorithm available in our module for {\it ab initio}
structure evolution ({\small MAISE}) package \cite{maise}. The key
differences from real evolutionary structure searches are that we (i)
allow only a few steps in local conjugate-gradient optimizations; (ii)
apply mutation operations only (distortions of atomic positions and
lattice vectors); and (iii) use relatively short runs. As can be seen in
Fig. 2, these measures have been found to promote sampling of
low-energy structures while our previously developed elimination of
similar structures based on radial distribution functions
\cite{ak16,ak23} maintained diversity in the evolved populations. Both
the low-accuracy conjugate-gradient relaxations and the high-accuracy
single-point calculations were done at the DFT level. Effectively, the
evolutionary sampling protocol is close to a Monte-Carlo-type
algorithm where the fitness factor plays the role of the proposal
acceptance distribution and where the automated generation of random
initial structures helps avoid biases.

For elemental systems we used unit cells with 1-6 atoms; two randomly
initialized populations of 16 structures were evolved over 20
generations. For binary and ternary systems we selected different
compositions with up to 10 and 6 atoms per unit cells,
respectively. After discarding 10\% of the highest-energy structures in
each run we obtained about 8,500, 7,500, and 9,000 structures for
unary, binary, and ternary systems, respectively. Table I summarizes
the number of data and the standard deviation in each
dataset. Figure 2(c) shows a wider dispersion of interatomic distances
in the generated set compared to the MD sets with larger but fixed
unit cells. The variable-cell evolutionary sampling eliminates the
artificial sharp feature and spreads out peaks from further neighbor
shells seen in the MD profile even at temperatures above melting.

\begin{table}[hb]
\hfill{}
\begin{tabular}{ll|cccccccccccc}\hline\hline
&Dataset &  & \multicolumn{3}{c}{stratified NN}      & \multicolumn{3}{c}{full NN}        \\
&        &  & $\Delta E$&  $\#$ of   &  $\#$ of      & $\Delta E$&  $\#$ of   & $\#$ of   \\
&        &  & eV/atom   &   data     &  weights      & eV/atom   &    data    &  weights  \\ \hline
& Cu     &  & 0.5172    & 8551       & 431           & --        & --         & 431          \\
& Pd     &  & 0.6262    & 8487       & 431           & --        & --         & 431          \\
& Ag     &  & 0.4481    & 8533       & 431           & --        & --         & 431          \\
& CuPd   &  & 0.5975    & 7623       & 1040          & 0.5829    & 24661      & 1902      \\
& CuAg   &  & 0.5391    & 7617       & 1040          & 0.6601    & 24701      & 1902      \\
& PdAg   &  & 0.5989    & 7601       & 1040          & 0.6833    & 24621      & 1902      \\
& CuPdAg &  & 0.2170    & 8917       & 660           & 0.6227    & 57329      & 5073      \\ \hline \hline
\end{tabular}
\hfill{}
\caption{A breakdown of datasets and NN adjustable parameters used in
  stratified and full training. Standard deviations ($\Delta E$) were
found individually for each unit cell size and then averaged over the
specified dataset. For example, the full CuPd dataset comprises all Cu
structures with 1-6 atoms, all Pd structures with 1-6 atoms, and all
Cu-Pd structures with 2-10 atoms; the CuPd dataset for stratified
training comprises only the binary structures.}
\end{table}

It should be acknowledged that the use of small-atom cells invariably
imposes certain constraints. For example, atomic environments in
1-atom structures are defined by only six parameters. Our following
tests indicate no apparent significant consequences for the
description of properties in the sampled regions from the presence of
the implicit biases. As discussed below in Sec. VI, one issue
detected in structure optimization runs with our trained NNs was an
occasional occurrence of short-distance configurations when the
starting structure already had unphysically short distances. The
strategy of adding such local minima to DFT datasets was found to be
ineffective, as artificial minima kept occurring even after a few
rounds of NN retraining. We found that complementing the original
dataset with a few short- and long-distance configurations (elemental
fcc, bcc, hcp, sc, dimer, and square as well as binary L$_{12}$, B2,
and dimer equilibrium structures scaled by up to $\pm30$\%) alleviated
the problem while having little influence on the NN overall
performance (Sec. VI). Discarding starting configurations with
interatomic distances below 70\% of the equilibrium ones ensured
robustness of NN-based evolutionary ground state searches.

\section{Atomic environment descriptors}

The description of interatomic interactions with both classical
potentials and NNs begins with breaking down the total energy into
atomic energies which defines the models' linear scaling with system
size. In the case of NNs, the representation of atomic environments
has to satisfy the following critical constraints: the NN input needs
to have a constant number of components and be invariant to
translations, rotations, and identical atom
swaps \cite{PhysRevB.87.184115}. The first constraint is a considerable
challenge because the number of nearest neighbors within a typical
6-\AA\ sphere can change, for instance, from over 70 in close packed
bulk structures to about half of that in surface geometries. Elegant
solutions for parsing arbitrary chemical environments into NN input
vectors include the smooth overlap of atomic positions (SOAP)
 \cite{PhysRevB.87.184115} and Parrinello-Behler (PB) symmetry
functions \cite{PhysRevLett.98.146401}.

In this study we have tested and used sets of pair ($G^1$) and triplet
($G^2$) PB symmetry functions:

\begin{eqnarray}
  G_i^1 = \sum_{j\ne i}^{all} e^{-\eta(R_{ij}-R_s)^2}f_c(R_{ij}), \\
  G_i^2 = 2^{1-\zeta} \sum_{j,k\ne i}^{all}(1+\lambda\cos\theta_{ijk})^\zeta\times \\ \nonumber
          e^{-\eta(R^2_{ij}+R^2_{ik}+R^2_{jk})}f_c(R_{ij})f_c(R_{ik})f_c(R_{jk}).
\end{eqnarray}

\begin{figure*}[t!]
\begin{center}
\includegraphics[width=140mm,angle=0]{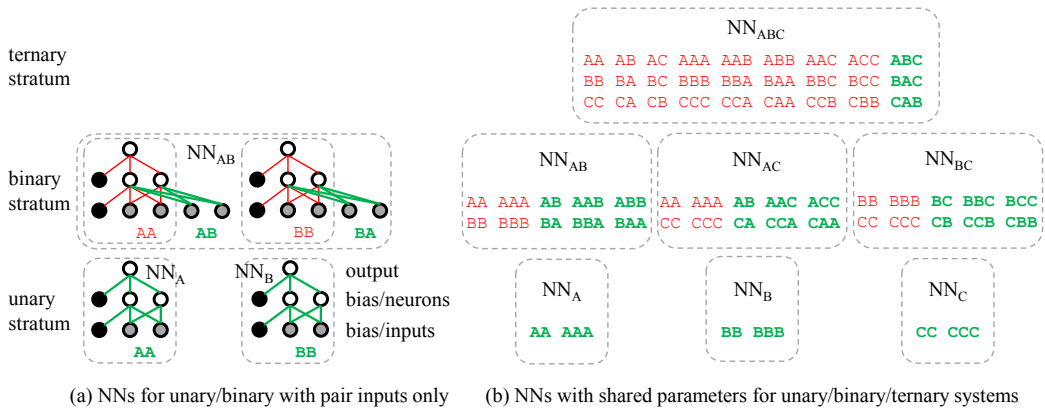}
\caption{ (color online) Schematic representation of stratified
  training of multilayer NNs. In (a) we show only one hidden layer of
  neurons and only pair input symmetry functions. Weights and element
  species shown in bold green are adjusted while the ones in thin red
  are kept fixed during the training from the bottom up. The letters
  denote species types in the input vector; for example, AAB describes
  a triplet symmetry function centered on an A-type atom with A-type
  and B-type neighbors (the order of neighbors is irrelevant).}
\vspace{-5mm}
\end{center}
\end{figure*}

The cutoff function $f_c(R_{ij})$ is $0.5\times(\cos(\pi
R_{ij}/R_c)+1)$ for $R_{ij}\le R_c$ and 0 otherwise. The explicit
values for $R_c$, $R_s$, $\eta$, $\zeta$ and $\lambda$  parameters are given in
the Supplementary Material \cite{suppmat}. We have found little sensitivity of the
resulting errors to the $R_c$ values in the 5-7 \AA\ range for the
Cu-Pd-Ag systems. The tests discussed in Section VI indicate that the
previously used set of 30 PB symmetry functions provides a fairly good
description of considered geometries while keeping the number of input
components reasonably low. The latter feature is important for
description of mulitcomponent systems because the number of NN
adjustable weights is almost linearly dependent on the length of the
input vector.

\section{NN setup}

We have implemented two widely used NN architectures: radial basis
functions (RBF), and multilayer perceptron (MLP). The former was shown
previously to provide a good description of many-body effects for
near-equilibrium graphite configurations  \cite{ak00}. However, RBF
sets with up to 1,000 Gaussians functions have shown inferior
performance for arbitrary geometries in elemental Cu, Pd, and Ag
systems. All NN results for multicomponent systems reported in this
study have been obtained with two-layer perceptrons.

For an atomic environment of atom $i$, the $N_0$ symmetry functions
$G_{n_0}^i$ are first rescaled into $x_{n_0}^i$ components of the
input vector as discussed below. The inputs and one bias $b_0$ are
then weighted, summed up, and transformed via nonlinear activation
functions $f(x)=\tanh(x)$ by $N_1$ first-layer neurons:

\begin{eqnarray}
  x_{n_1}^i = f\left[\sum_{n_0}^{N_0}w_{n_0,n_1}^{(0)}x_{n_0}^i + b_{n_1}^{(0)}\right]. \nonumber
\end{eqnarray}

The neuron outputs and the first-layer bias $b_1$ are processed in a
similar way by $N_2$ second-layer neurons:
\begin{eqnarray}
  x_{n_2}^i = f\left [\sum_{n_1}^{N_1}w_{n_1,n_2}^{(1)}x_{n_1}^i + b_{n_2}^{(1)}\right ]. \nonumber
\end{eqnarray}

Finally, the resulting values are added up to be the NN energy output:
\begin{eqnarray}
  E_i = \sum_{n_2}^{N_2}w_{n_2}^{(2)}x_{n_2}^i + b^{(2)}. \nonumber
\end{eqnarray}

The standard way of training such NNs involves the adjustment of
weights and biases to minimize the squares of the errors between
target and NN-produced values \cite{bishop}. The known complication in
training NN models is that only the {\it total} rather than {\it
  atomic} energies are available in quantum-mechanical calculations
\cite{ak00,PhysRevB.90.104108,PhysRevLett.98.146401}. The local
information can be introduced in the form of the atomic forces but
they do not define the last-layer biases and are more prone to
numerical noise. Both energy- and force-based fittings have been
discussed and used in previous studies
\cite{ak00,1.3095491,extrabad,PhysRevB.85.045439}. We have implemented
analytical derivatives for each fitting type and rely on either
Broyden-Fletcher-Goldfarb-Shanno (BFGS) or conjugate-gradient
minimizer to drive the least squares optimization. All presented
results have been obtained with total energies as target values.

A preventive measure against favoring certain input components is to
determine the full ranges of the symmetry functions in the training
set and rescale them to be between --1 and 1. It is a good practice to
start with NN weights that make neurons' inputs have the average of 0
and the standard deviation of 1 in {\it all} layers because very large
initial values of $x$ result in small derivative values of the
$\tanh(x)$ activation function and, ultimately, in a slow or
suboptimal weight convergence  \cite{initialize}. We have adopted a
common approach for random initialization of weights ensuring good
starting values for all $x$.

\section{Stratified training for multicomponent systems}

NNs can be and have been successfully used for modeling multicomponent
systems but it should be kept in mind that the total number of
parameters grows rapidly with the number of elements (see Table
I). The formalism generalization requires an extension of symmetry
function definitions to multiple element types, specification of NN
links to new input components, selection of the training procedure,
etc. In a previous study by Artrith {\it et al.} \cite{PSSB:PSSB201248370},
a ternary NN for Cu-ZnO trained on a combined set of unary, binary,
and ternary data displayed a good performance for elemental systems
\cite{PSSB:PSSB201248370}. The finding is encouraging but NNs for
new multicomponent systems would need to be constructed from scratch
and would give close but not identical values for shared subsets of
elements. 

The central question investigated in this study is whether
multicomponent NNs could be built from the bottom up and whether they
would still be accurate. The procedure shown in Fig. 1 illustrates how
separate datasets can be used to optimize the corresponding weights in
a natural sequential order. For example, AB binary parameters are
trained to the AB database only after A and B elemental parameters are
fitted to the A and B databases. Such a hierarchical approach to
fitting parameters has been previously used to build semi-empirical
models, such as Gupta-type potentials \cite{jp1048088,JCC:JCC21197},
(M)EAM-level potentials \cite{Luyten20071668,Luyten2009370}, and
tight-binding models \cite{PhysRevB.84.155119}. Here we examine the
applicability of the stratified fitting strategy in the context of
NNs.

The implementation of the scheme is fairly straightforward provided
that two points are handled properly. The first one concerns defining
the ranges for interspecies symmetry functions. One needs to ensure
that when A-type atoms have no B-type neighbors the NN 'AB' input
components $x_{n_0}^i$ are strictly 0. However, the minimum value
$G_{n_0}^{min}$ for an AB symmetry function in the full AB binary
database might not be 0 because all atoms of species A happened to
have at least one neighbor of species B. Then even for a purely A-type
structure the non-zero AB-type
$x_{n_0}^i=(0-G_{n_0}^{min})/(G_{n_0}^{max}-G_{n_0}^{min})$ will
produce an artificial contribution to the total energy. In all
generated databases the $G_{n_0}^{min}/G_{n_0}^{max}$ ratios turned
out to not exceed 0.4 and a simple solution to set all $G_{n_0}^{min}$
values to 0 has been found to work well.

The second point concerns specifying which subsets of parameters are
allowed to vary in the stratified training. One needs to ensure that a
multicomponent NN fitted to the corresponding database produces
unchanged output values for constituent elemental/binary structures.
Fig. 3(a) helps illustrate that the requirement is met if only the
weights connected to the interspecies inputs are allowed to adjust.
The input bias weights, in particular, should not be changed because
they define the reference energy of a free atom under the previously
discussed condition of $G_{n_0}^{min}=0$ for all $n_0=1,\dots,N_0$.

The following benchmark results illustrate the performance of NNs
trained in the standard and stratified fashions. The tests were
designed to allow comparisons for practical sizes of datasets and NN
parameters that can be generated in future studies for a large set of
systems.

\section{Results}

\subsection {Total energy errors}

\begin{table*}[t!]
\hfill{}
\begin{tabular}{ll|llllllllllll|llllllllllll|lllllllll}\hline\hline
&             &   &\multicolumn{2}{c}{Cu}&&&\multicolumn{2}{c}{Pd}&&&\multicolumn{2}{c}{Ag}&&&\multicolumn{2}{c}{CuPd}&&&\multicolumn{2}{c}{CuAg}&&&\multicolumn{2}{c}{PdAg}&&&\multicolumn{2}{c}{CuPdAg} \\
&             &   &train    &test        &&&train    &test        &&&train    &test        &&&train    &test          &&&train    &test          &&&train    &test          &&&train    &test   \\  \hline\hline
& NN$^{\text{strat.}}$  &   &3.88     &3.96        &&&10.49    &10.91       &&&4.08     &3.92        &&&7.06     &7.37          &&&3.56     &3.84          &&&5.52     &6.21          &&&6.45     &6.28    \\
& NN$_{\text{CuPd}}^{\text{full}}$& &4.72 &4.85        &&&11.16    &10.84       &&&--       &--          &&&7.07     &7.67          &&&--       &--            &&&--       &--            &&&--       &--      \\
& NN$_{\text{CuAg}}^{\text{full}}$ &&4.63 &4.84        &&&--       &--          &&&4.56     &4.48        &&&--       &--            &&&3.91     &4.13          &&&--       &--            &&&--       &--      \\
& NN$_{\text{PdAg}}^{\text{full}}$ &&--   &--          &&&11.06    &11.25       &&&4.79     &4.63        &&&--       &--            &&&--       &--            &&&5.91     &6.05          &&&--       &--      \\
& NN$_{\text{CuPdAg}}^{\text{full}}$ &&5.24 &5.27   &&&11.32 &11.45  &&&4.97 &4.87        &&&7.83     &8.12          &&&4.31     &4.43          &&&6.12     &6.57          &&&6.36     &6.70    \\ \hline\hline
\end{tabular}
\hfill{}
\caption{Training and testing errors in meV/atom for different
  systems, evaluated by NNs trained in full and stratified fashions.}
\end{table*}

\begin{figure}[b!]
\begin{center}
\includegraphics[width=80mm,angle=0]{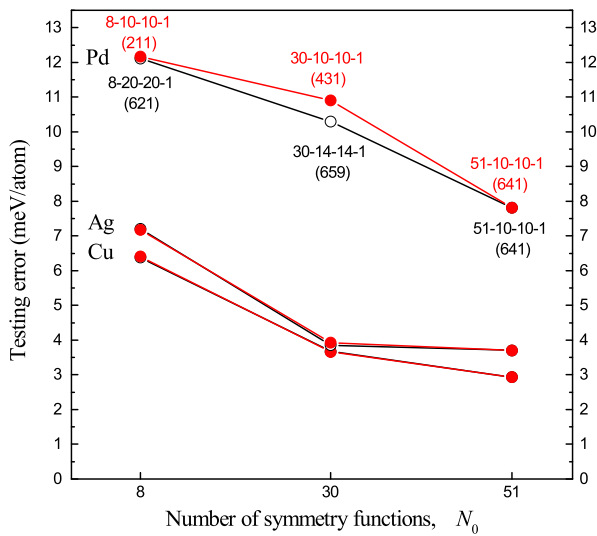}
\caption{(Color online) Testing error as a function of number of
  symmetry functions and NN parameters. The red solid points
  correspond to $N_0$-10-10-1 NNs while the black open points
  correspond to NNs with varying numbers of neurons chosen to be close
  to the total number of adjustable parameters for different $N_0$. A
  noticeable decrease in the testing error is seen only for Pd and
  $N_0=30$. }
\end{center}
\end{figure}

In all considered cases we have kept the data to parameter ratio above
7:1 to avoid overfitting. We found the residual and testing errors to
differ by less than 5\% in cross validation or random selection tests
with 10\% of data allocated for this purpose. In order to select a
suitable descriptor size and a sufficient number of NN parameters
for the analysis of multicomponent systems we first examined the
dependence of the testing error on these settings for the Cu, Pd, and
Ag elemental systems.

Figure 4 shows the accuracy of the total energy description using
$N_0$-$N_1$-$N_2$-$1$ NNs with $N_0= $8, 30, or 51 PB symmetry
functions and $N_1=N_2= $10, 14, or 20 neurons. The 30 and 51 sets have
been used previously in the investigation of elemental systems
\cite{PhysRevB.85.045439}. We also considered a subset of only eight
$G^{1}$-type pair symmetry functions to illustrate the importance of
many-body interactions in these systems. For NNs with the same number
of neurons, $N_1=N_2=10$, the inclusion of $G^{2}$-type triplet
symmetry functions led to substantial drops in the error for Cu, Pd,
and Ag by 43\%, 10\%, and 45\% at 30 components and another 20\%,
28\%, and 5\% at 51 components.

Since variations in the descriptor size affect the number of NN
adjustable parameters, we repeated the tests for 8 and 30 components
using $N_1=N_2=20$ and $N_1=N_2=14$ components, respectively. The
black points in Fig. 4 correspond to the NNs with fairly matched total
numbers of free parameters. The little improvement over the
$N_0$-$10$-$10$-$1$ results reveals that the accuracy is defined
predominantly by the completeness of the symmetry function sets for
these systems.

The considerably higher errors for Pd are only partially explained by
the larger standard deviations in the generated datasets (Table I).
Given that the PB symmetry functions are fairly general and that
Pd and Ag are of similar size, the different NN performance likely
arises from the different complexities of the potential energy
surfaces. Our preliminary tests carried out with the same settings for
neighboring elements in the periodic table suggest that the accuracy
observed for Pd is more of a rule rather than an exception.

Overall, the 30-10-10-1 choice for elemental NNs ensures reasonably
good accuracy and keeps the number of parameters relatively
low. Non-elemental systems were consequently modeled with 82-10-10-1
binary and 156-10-10-1 ternary NNs. For benchmarking purposes we first
employed the standard training approach to generate all combinations
of elemental, binary, and ternary NNs for Cu, Pd, and Ag (we use
NN$^{\text{full}}_{\text{AB}}$ and NN$^{\text{full}}_{\text{ABC}}$
notations to refer to NNs with the full sets of parameters adjusted in
the binary and ternary cases, respectively). For example, for the CuPd
binary we used 8551 Cu, 8487 Pd, and 7632 CuPd structures and
optimized all 1902 parameters in the 82-10-10-1 NN. Merging sets with
similar data/parameter ratios should keep the average errors close but
the transfer of errors could be a cause for concern. Namely, in the
minimization of the total error function the Cu-Cu weights are defined
not only by Cu environments in Cu and Cu-Pd structures but also,
indirectly, by Pd environments in Cu-Pd and Pd structures.

\begin{figure*}[t!]
\begin{center}
\includegraphics[width=175mm,angle=0]{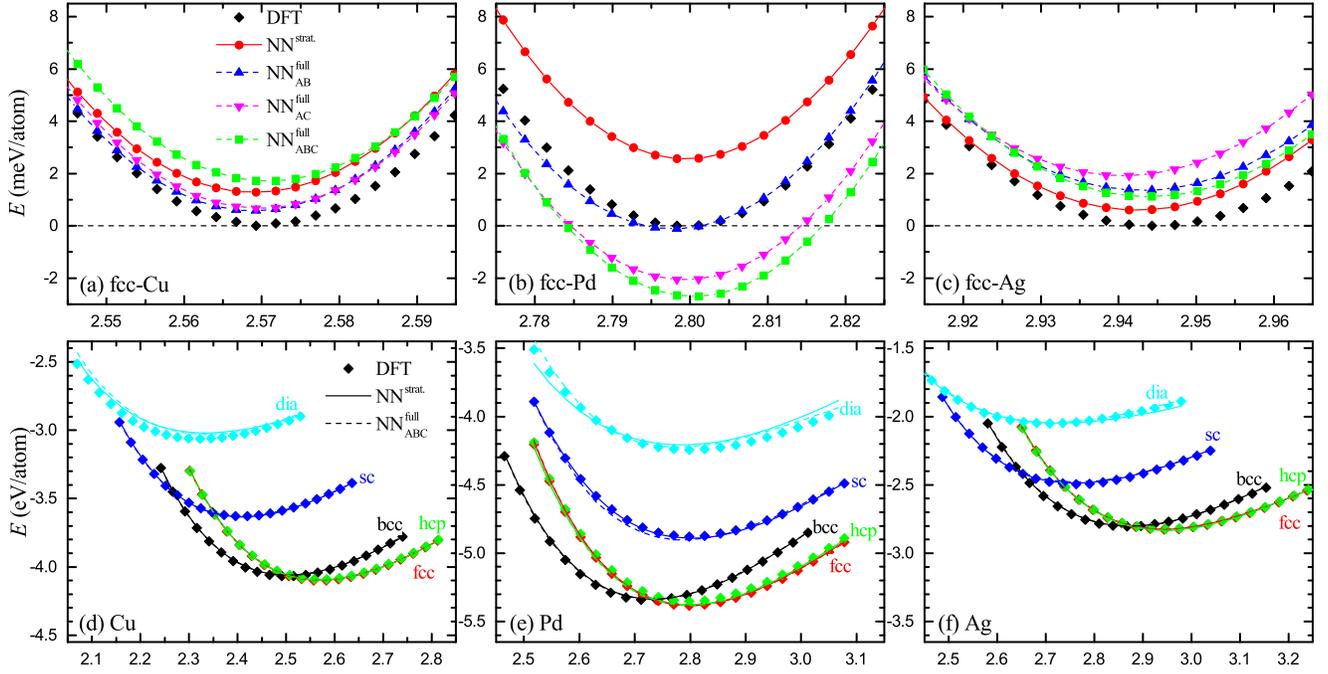}
\caption{Total energy as a function of nearest neighbor distance
  evaluated with DFT and NNs for Cu, Pd and Ag. The top panels show
  data for the fcc ground states near equilibrium; the colors and
  types of connected symbols denote the NN flavors. The bottom panels
  show results for different structures in a wider nearest neighbor
  distance range; here the solid and dashed lines denote the
  stratified and full ternary NNs while the colors and types of
  symbols mark structure types.}
\end{center}
\end{figure*}

It has been shown previously that a full training of the Cu-ZnO
ternary NN produced a reasonably good elemental Cu model
 \cite{PSSB:PSSB201248370}. Systematic tests summarized in Table II
indicate that the error transfer is fairly small for the subsystems in
our multicomponent NNs as well. For instance, the largest increase in
the testing error, from 3.96 meV/atom with NN$^{\text{strat.}}$ up to
5.27 meV/atom with NN$^{\text{full}}_{\text{CuPdAg}}$, is observed for
Cu since the model accuracy for this element is influenced in the full
training by the lower accuracy of the Pd and Ag descriptions. Judging
by the testing errors across all considered systems, the performance
of the stratified NNs is comparable to or better than that of the
standard NNs.

Figure 5 details how the NNs constructed in different fashions fair
against each other and the DFT data for select high-symmetry
structures. Our data generation protocol favored the sampling of
configurations around low-energy states (see Fig. 2). Hence, it is not
unexpected to see good description, within 2-3 meV/atom, of the fcc
ground states in Fig. 5(a-c) which is important for calculation of
defect energies and phase diagrams. On a larger scale shown in panels
(d-f), it is satisfying to see a good behavior of all NNs even for the
high-energy diamond configurations unnatural for these metals. Fig.
S1 demonstrates that all these structures are described very well
with NNs trained without the bcc, fcc, hcp, and sc equation of state
data. In fact, the extra set included to sample short-distance
configurations caused a shift away from the ground state energy around
the equilibrium distance.

\subsection {Defect formation energies}

\begin{figure*}[t!]
\begin{center}
\vspace{0.0 cm} \includegraphics[width=175mm]{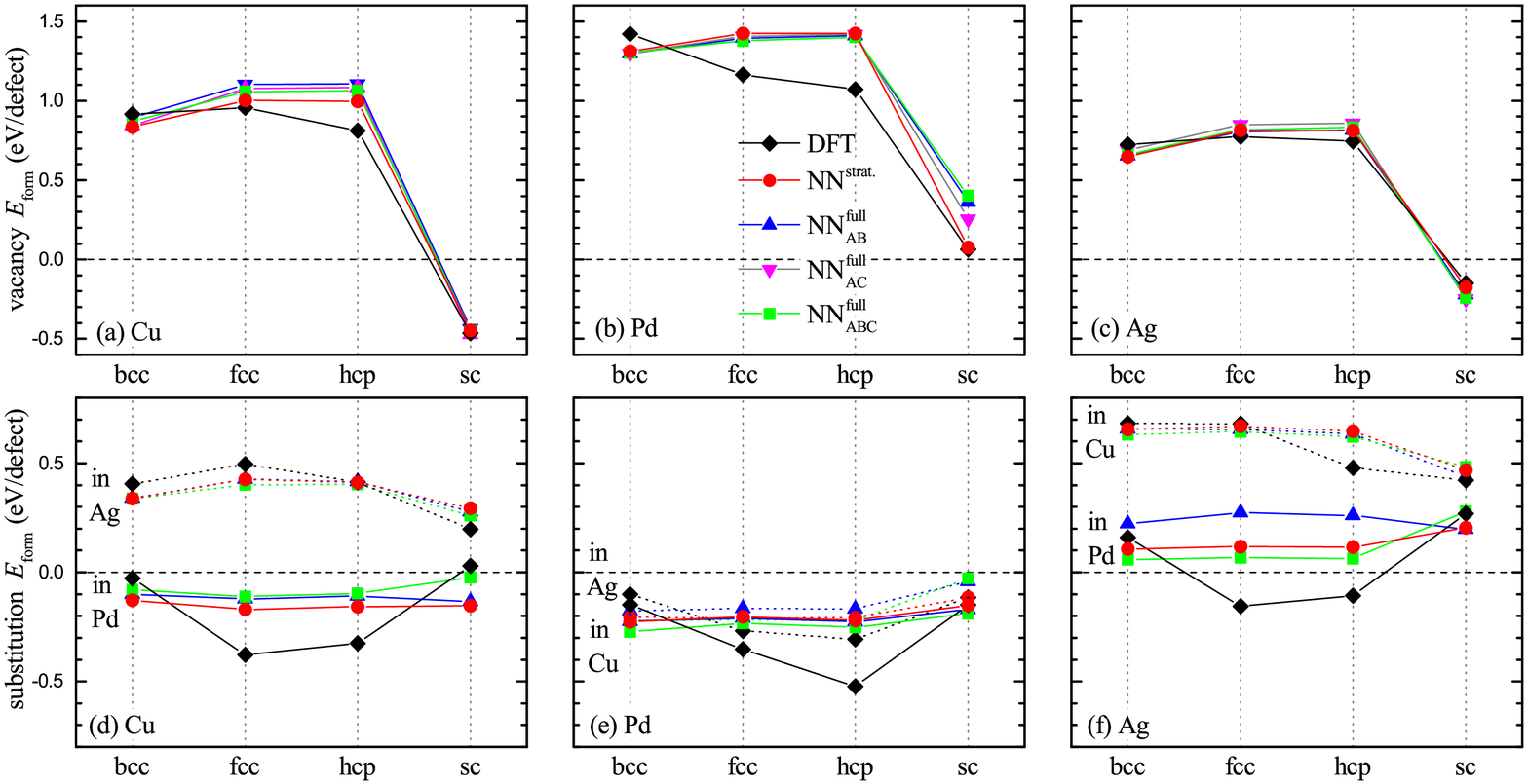}
\caption{Defect formation energies in high-symmetry Cu, Pd, and Ag
  structures evaluated with DFT and different NNs. The top and bottom
  panels correspond to vacancy and substitutional defects,
  respectively. In (d-f), the solid and dotted lines are used to
  differentiate into which ideal elemental lattice an atom is
  embedded; e.g., in (d) Cu replaces either Ag or Pd atoms.}
\end{center}
\end{figure*}

Evaluation of defect energies is one of the most revealing tests of NN
performance because the unit cells generated and used for training are
not sufficiently big to simulate defect environments. Vacancy and
substitutional defects were calculated in medium-sized bcc, fcc, hcp,
and sc supercells. As common in such comparisons, only the atomic
positions in the supercells were optimized at the DFT level while the
lattice constants were kept fixed at DFT-optimized values. The vacancy
formation energies of 0.96, 1.16, and 0.77 eV/vacancy evaluated with
DFT in $3\times 3\times 3$ 26-atom fcc supercells for Cu, Pd, and Ag
were close to respective previously reported DFT values of 0.99, 1.21,
and 0.73 eV/vacancy \cite{vacancy}. The NN energies, as in all our
tests, were compared for the final structure optimized with DFT but we
also checked that following atomic relaxations with NNs resulted in
energy gains less than 0.01 eV/defect.

Figure 6 summarizes the defect formation energy results for all
constructed NNs. It can be seen that the type of training had little
effect on how the NNs reproduce the DFT values. The accuracy of all
Cu-Ag NNs is fairly good in these tests, typically within 10-30\% of
$E^{\text{defect}}_{\text{form}}$. For the defect structures involving
Pd, the NNs display noticeable deviations reaching 0.4 eV/defect.
Surprisingly, the largest errors occurred when the substitution of Ag
for Pd in the fcc matrix is evaluated with
NN$_{\text{PdAg}}^{\text{full}}$ specifically tuned to this binary.
 The 0.2-0.4 eV/defect discrepancies are not unexpected
  because these values, calculated per {\it unit cells} of about 30
  atoms, correspond to $\sim10$ meV/atom errors. It is still worth
  examining how different factors affect the NN performance in this
  case.

A widely discussed issue is that the scope of NN applicability extends
as far as the dataset does. In particular, simulation of structures
with any of the symmetry function values outside of the sampled range
corresponds to the NN extrapolation regime and is not expected to
produce reliable outputs \cite{extrabad}. Fig. 7 illustrates that
creation of a vacancy in the $3\times 3\times 3$ 27-atom fcc-Pd causes
small deviations of atomic environments from the ideal one when
compared to the distribution of input values in the full elemental Pd
dataset. However, even if all input components of new structures to be
evaluated are between the corresponding minimum and maximum values of
the training set the points could be far away from any data clusters
in the large $N_0$-dimensional configuration space. 

Addition of the Ag-in-fcc-Pd defect structure into the
  training set reduced the 0.27 eV/defect discrepancy between DFT and
  NN$^{\text{strat.}}$ down to 0.25, 0.18, and 0.02 eV/defect if the
  data point was weighted by a factor of 10, 100, and 1000,
  respectively. For comparison, increasing the number of descriptor
  components to 51 reduced the error down to 0.22 eV/defect. Clearly,
  describing the defect energy better than 0.1 eV/defect is difficult
  given the $\sim$4-7 meV/atom overall accuracy of the constructed
  NNs.

\begin{figure}[b!]
\begin{center}
\vspace{0.0 cm}
\includegraphics[width=80mm]{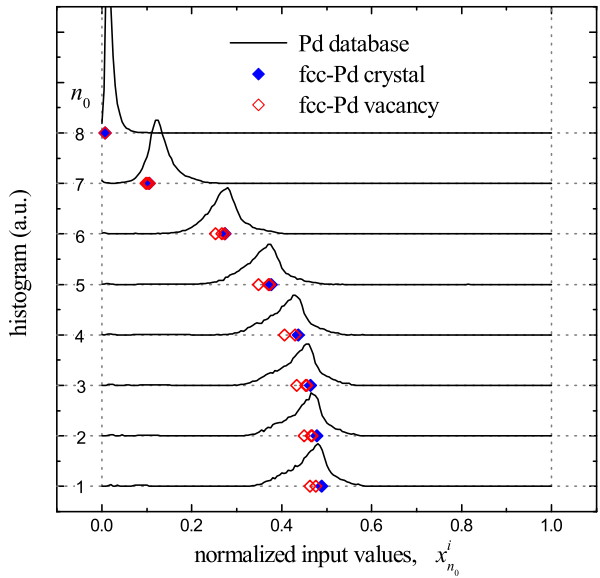}
\caption{Distribution of normalized input components in the full Pd
  database and select structures. Creation of a vacancy in the
  $3\times 3\times 3$ 27-atom fcc-Pd supercell results in 3 Wyckoff
  sites. The input values corresponding to their atomic environments
  are close to the ideal ones in fcc-Pd and to the middle of
  distribution peaks (only the first 8 pair-symmetry function values
  are shown for clarity). The sites contribute to
  $E^{\text{vac}}_{\text{form}}=1.42$ eV as follows:
  $12(0.210)+8(-0.072)+6(-0.087)$ eV. }
\end{center}
\end{figure}

\subsection {Compound formation energies}

\begin{figure}[t!]
\begin{center}
\vspace{0.0 cm}
\includegraphics[width=87mm]{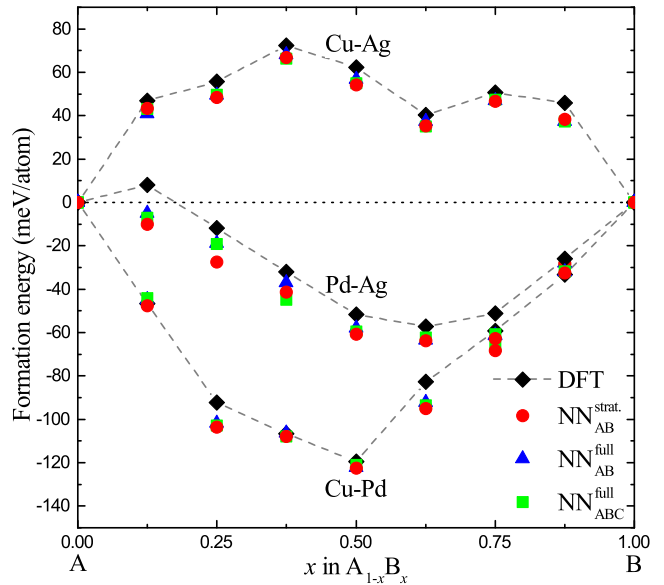}
\caption{Formation energies calculated with NN and DFT methods for
  lowest-energy 8-atom binary structures found in evolutionary
  searches based on NN$^{\text{strat.}}$ interatomic models. The
  dotted lines are a guide for the eye.}
\end{center}
\end{figure}

Accurate evaluation of compound formation energies is a critical
benchmark for the developed models to be used for materials
prediction. The sign determines the miscibility of elements while the
magnitude determines which compositions are thermodynamically stable
via construction of the convex hull \cite{Curtarolo2005163}. The absolute values
of the formation energy in the considered binaries are relatively low,
about 100 meV/atom but high enough to be resolved with NNs.

\begin{table}[!b]
\hfill{}
\begin{tabular}{cc|ccccccccccc}\hline\hline
& binary   &\multicolumn{4}{c}{$\Delta E$ in meV/atom} & \\
&          & NN$^{\text{strat.}}$ &NN$_{\text{AB}}^{\text{full}}$    & NN$_{\text{ABC}}^{\text{full}}$ & DFT$_{\text{relaxed}}$     \\  \hline
&  CuPd    &     7.5        & 5.7      & 6.6   &  1.3  \\
&  CuAg    &     6.0        & 5.6      & 6.0   &  0.8  \\
&  PdAg    &    11.7        & 8.1      & 9.7   &  0.9  \\ \hline\hline
\end{tabular}
\hfill{}
\caption{Column (1) Binary systems investigated with unconstrained
  evolutionary searches as described in the text. Columns (2 and 3)
  Formation energy errors in 3 sets of 8-atom binary structures
  identified in evolutionary searches (see Fig. 7). Column (4) Average
  gain in energy after local relaxations of NN-optimized structures
  with DFT.}
\end{table}

For our first test, we considered fcc and bcc solid
  solutions common in these binaries. As described in the
  Supplementary Material \cite{suppmat}, we simulated the alloys in
  known special quasirandom structures \cite{SQS,SQSbcc,SQSfcc} and
  in our own sets of 8- and 16-atom random structures. The average
  errors for these phases were lower than the corresponding NN errors
  by 1-2 meV/atom in nearly all cases which can be expected for the
  well-sampled low-energy part of the configurational space
  (Fig. S2).

In order to generate a more diverse test set of binary
structures and check the robustness of the trained NNs we performed a
series of NN-based evolutionary ground searches. For the three
binaries, we considered A$_{8-n}$B$_n$ ($n=1\dots 7$) compositions and
used a typical set of evolutionary settings from our previous
DFT-based searches \cite{ak16,ak23}. 16 members in populations were
evolved over 50 generations, as 70\% and 30\% of children were created
with crossover and mutation operations, respectively. The former
involved mixing of approximately halves of two parent structures and
the latter consisted of lattice distortions, atom displacements, and
atom swaps. The most stable structure was typically found within the
first 10 generations. The number of structures detected to have short
interatomic distances at the beginning of or during local
optimizations was small, 0.4\%, 0.1\% and 4.4\% for CuPd, CuAg, and
PdAg, respectively.

An important outcome of these simulations summarized in Fig. 8 is that
the developed interpolation-type models with numerous parameters
showed no clearly unphysical low-energy configurations. A similar
behavior was observed in evolutionary runs for the elemental and
ternary systems (not shown). The formation energy differences
calculated with DFT and NN for the sets of seven binary lowest-energy
structures in Table III correlate well with the NNs testing errors in
Table II. Note that because $E_{\text{form}}(A_{1-x}B_x)$ is defined
as $E(A_{1-x}B_x)-(1-x)E(A)-xE(B)$ the reference energy errors for the
elements add to the total formation energy errors.
NN$^{\text{full}}_{\text{AB}}$ tuned to the corresponding binaries are
in slightly better agreement with DFT than NN$^{\text{strat.}}$ or
NN$^{\text{full}}_{\text{ABC}}$.

The largest deviations for all three NNs happened in the Pd-Ag binary
at the Pd-rich end, the composition limit with the least accurate
description of the substitutional defect energies in Fig. 6. Notably,
the NN results were consistently below the corresponding DFT values
rather than scatter around them. Because of this trend, relative
energies calculated needed for the construction of a convex hull
should be more accurate. Nevertheless, a reliable identification of
thermodynamically stable alloys at different stoichiometries requires
resolution of relative energies typically with 1-2 meV/atom accuracy
which is difficult to achieve. It is important, however, that the
relaxed crystal structures produced with the NN-based evolutionary
searches are very close to local minima in the DFT treatment, as
additional DFT-based relaxation lowers the energy by about only 1
meV/atom (Table III). Hence, NN-based searches can provide a pool of
viable candidates that can be quickly tested at the DFT level.

\begin{figure*}[t!]
\begin{center}
\vspace{0.0 cm}
\includegraphics[width=175mm]{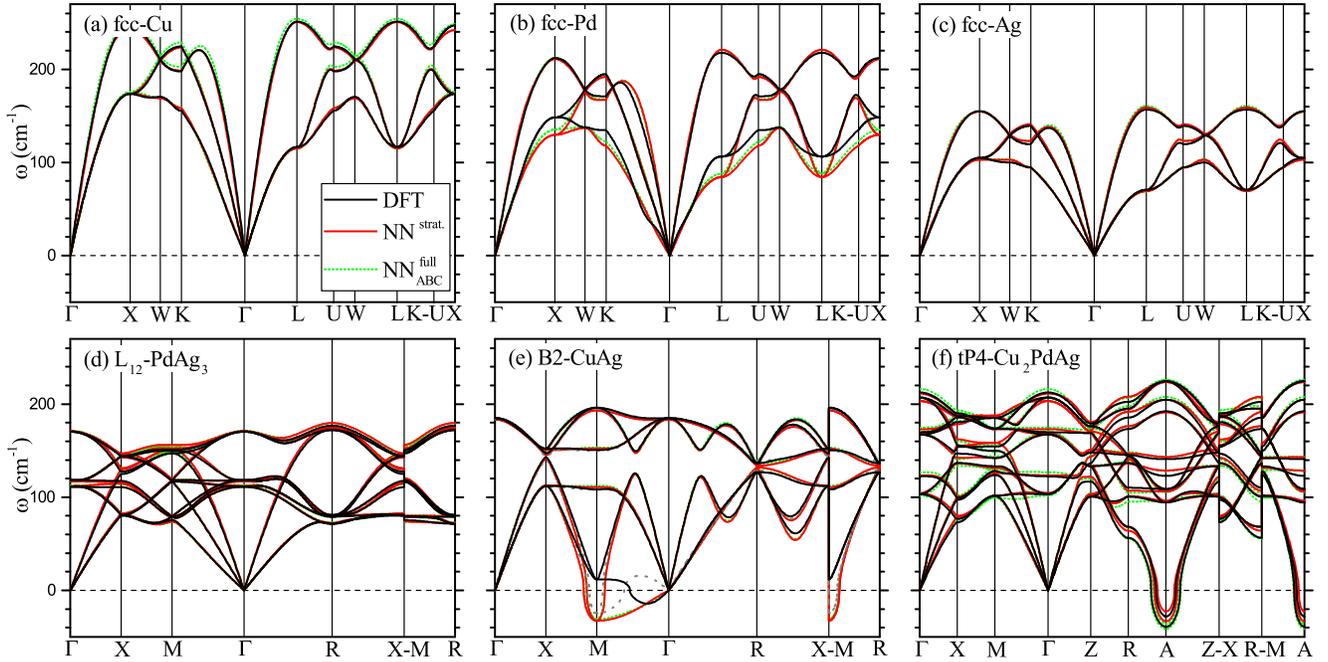}
\caption{Phonon dispersions calculated with the frozen phonon method
  at the DFT, stratified NN, and full NN levels. The dotted gray lines
  in (e) illustrate the change in the soft-mode frequencies calculated
  with DFT when the supercell size is reduced from $4\times 4\times 4$
  to $3\times 3\times 3$.}
\end{center}
\end{figure*}

\subsection {Phonons}

Evaluation of atomic-force defined quantities gives valuable
information about classical models because the derivatives depend on
neighbors twice the cutoff distance away from the considered atom
 \cite{ak00}. We used the frozen phonon method as implemented in PHON
 \cite{PHON} to calculate force constants and phonon dispersions for
representative structures. The same supercells (above 108 atoms) and
settings (displacement size 0.04 \AA) were used in DFT and NN
calculations. Dispersions were plotted along high-symmetry directions
as defined in Ref. \onlinecite{kpath}.

The top panels in Fig. 9 show phonon frequencies calculated with the
help of $6 \times 6\times 6$ expansions of primitive fcc cells using
DFT (black), stratified NN (red), and full NN (green) methods. We
observe essentially no difference between the results obtained with
the two types of NNs and an excellent agreement with the DFT data. As
in the other tests, the most noticeable deviations (for some optical
modes) are seen in the challenging Pd case. The acoustic modes are
reproduced well even for Pd which can be attributed to insignificance
of constant energy shifts, shown in Fig. 9, in the calculation of
derivatives. 

The bottom panels in Fig. 9 show phonons in multicomponent structures
based on fcc (d,f) and bcc (e). Again, the two NN types produce very
similar dispersions remarkably close to the DFT curves in (d) and
(f). It is satisfying to see that the NNs captured the large softening
of certain modes and the presence of imaginary frequencies shown as
negative values in (e) and (f). The discrepancy for the lowest
$M$-point frequencies in (e) is not surprising due to the known
sensitivity of phonon results to simulation settings at the DFT level
\cite{ak33,HSE-phon}. For instance, by reducing the supercell from
$4\times 4\times 4$ to $3\times 3\times 3$ in which the $M$-point is
not folded onto $\Gamma$ we obtained a very different shape of the
lowest phonon branch along $\Gamma-M$ with DFT; in contrast, the
dispersions calculated in the smaller supercell with the NNs were
essentially unchanged.  Further tests in the Supplementary
  Material \cite{suppmat} (Fig. S3) illustrate the difficulty of
  obtaining converged DFT results. Nevertheless, it is evident that
  the phonon softening magnitudes with respect to the 110 cm$^{-1}$
  unsoften mode frequency at the $M$ point in Fig. 9(e) are
  comparable in the two treatments: $\sim 100$ cm$^{-1}$ with the DFT
  and $\sim 140$ cm$^{-1}$ with the NN.  The observed ability of the
NNs to identify dramatically softened modes is crucial for
constructing dynamically, and possibly thermodynamically, stable
derivatives in structure searches.

\begin{table}[!b]
\hfill{}
\begin{tabular}{ll|ccccccc|ccccccccccccccccccccccccccccccccccccc}\hline\hline
& model                             &&\multicolumn{5}{c}{$\Delta F-\Delta F^{\text{DFT}}$ } &&&\multicolumn{5}{c}{$\Delta F_\text{form}$ } \\
&                                   &&Cu           & &Pd       &&Ag      &&&  CuAg     &&PdAg$_3$   &&Cu$_2$PdAg   \\  \hline
&DFT                                &&  -          & & -       && -      &&& -21.0     &&6.5        &&-10.9        \\
&NN$^{\text{strat.}}$               && 0.3         & & -8.7    && 2.8    &&& -27.9     &&9.3        && -8.8        \\
&NN$_{\text{CuPdAg}}^{\text{full}}$ && 2.9         & & -3.9    && 4.3    &&& -28.2     &&6.4        && -9.5        \\ \hline\hline
\end{tabular}
\hfill{}
\caption{Thermal corrections to free energy (elemental sytems) and formation free energy (multielement systems) at 1000 K in meV/atom.}
\end{table}

Another important use of NNs is inclusion of vibrational entropy
contributions to the free energy at finite temperatures
 \cite{PHON}. Given the generally good agreement for the phonon modes
along the high-symmetry directions, one could expect a reasonably
accurate estimate of the phonon DOS integrated over the full Brillouin
zone as well. Using a standard expression $\Delta F =
k_BT\int_0^{\infty}d\omega g(\omega) \ln(2\sinh(\hbar\omega)/2k_BT)$
 \cite{PHON}, we calculated the quasiharmonic corrections at 0 and 1000
K for the 6 considered structures \cite{comment}. The zero point energies, found to be
between 20-30 meV/atom in magnitude, were described by the NNs within
0.6 meV/atom of the DFT values for all 6 phases. The corrections went
up to 400-500 meV/atom at 1000 K but were still described well with
NNs. Table IV illustrates that the mismatches between NN and DFT
results remained below 9 meV/atom for the elements. The NNs captured
the sign and the magnitude, within 7 meV/atom, of the vibrational
contributions to the {\it formation} free energies for the 3
compounds. The tests suggest that the NN models linearly scaling with
system size can be used for a fast examination of compound stability
at realistic synthesis conditions.

\section {Discussion and conclusions}

Having presented our main results and observations, we expand on our
discussion of the NN model robustness, parameter meaning, expected
advantages/limitations, and applicability scope.

Development of a robust training procedure requires resulting NNs be
not sensitive to the particular weight initialization and choice of
the data subsets. During the course of this work, we have constructed
over 500 various Cu-Pd-Ag NNs and observed little dependence of the
model performance on these factors. Namely, after 60,000 steps of BFGS
or conjugate-gradient optimization, the residual and testing errors invariably
converged to values within 3-5\% of each other. The close accuracy of
NNs trained to different data subsets suggest that there are many
comparable local minima in the large configuration space of NN
parameters. The observation is supported by our additional tests in
which we monitored various NN outputs as the models were being
tuned. As shown in Fig. S4, we calculated formation energies, phonon
frequency, etc., using three differently initialized NNs after every
20,000 training steps. After 60,000 steps, all probed properties
converged to very close values in all three cases. It might still be
worth trying more advanced error minimization techniques, such as the
evolutionary algorithm, but so far we have not observed a particular
need for that.

Interpretation of the NN inner workings is notoriously difficult due
to the sheer quantity and the hierarchical connectivity of adjustable
weights. We looked for possible parameter correlations in simplified
one-layer 16-10-1 binary NNs with eight pair symmetry functions per
element. As can be seen in Fig. 3(a), the weights connecting the
inputs and the first-layer neurons define the interspecies interaction
strength. Suppose all $w^{(0),\text{AA}}_{n_0,n1}$ and
$w^{(0),\text{AB}}_{n_0,n1}$ pairs for every symmetry function $n_0$
were identical; then a substitution of A for B would not change the
energies of any neighboring A-type atoms. We trained full and
stratified NNs for CuPd, CuAg, and PdAg using 3 different data subsets
for each binary (18 NNs in total). Typical distributions of
$w^{(0),\text{AA}}_{n_0,n1}$ and $w^{(0),\text{AB}}_{n_0,n1}$ weights
plotted in Fig. S5(a,b) do not exhibit particularly strong
correlations. After fitting all 36 sets with linear dependences and
plotting Pearson's $r$ values in Fig. S5(c) we observed a consistently
better correlation of weights in NNs trained in the full fashion,
i.e., when AA, AB, BB, BA, and all biases were adjusted
simultaneously. Fitted linear slopes were found to be all positive,
with angles between 10 and 80 degrees for all trained NNs which
indicates an overall tendency for AB and AA parameters to have the
same sign. Ideally, one would want to break down the results further
by symmetry functions to establish which inputs are most significant
for defining interspecies interactions or how to initialize weights
for faster convergence. Unfortunately, even for these simplified NNs
we could not extract any other noticeable trends.

It is worth clarifying further how weight constraining affects the
flexibility of NNs to describe multicomponent systems. Fig. 3 shows
that there are no more adjustable parameters beyond the ternary set
and it may appear that NNs cannot be tuned to model quaternary systems
due to stratification of training.  In reality, the limitation is
imposed by the use of pair and triplet symmetry functions. Without
explicitly including quadruplet symmetry functions to account for
4-body interaction terms, the adjustment of all weights to the full
dataset with the standard training would not carry much physical
meaning. In this respect, matching data and parameter sets in the
stratified approach seems more transparent and justifiable.

The demonstrated good performance of both standard and stratified NNs
could be related to the similarity of the considered
elements. Treatment of systems comprised of elements with very
different bonding mechanisms, such as metal hydrides, oxides, etc., is
certainly more challenging  \cite{NNcharge}. With the top-level weights
fixed, the modulation of, e.g., metal AA inputs by metal-hydrogen AB
inputs before the net signal is fed into the first-layer neurons might
be insufficient to fully capture the hybridization and charge transfer
effects between the elements with such differences in size and
electronegativity. The actual limitation in both training schemes is
due to the adopted linkage of inputs to the first-level neurons
only. Since fully trained NN with this architecture have performed
well for Cu-ZnO and NaCl \cite{PSSB:PSSB201248370,PhysRevB.92.045131}
systems, one can expect a comparable performance for stratified NNs. A
way to relax the architecture-defined constraint for both fitting
schemes is to link the inputs to higher-level neurons as well. In
fact, generation of reliable DFT data is a greater challenge for
strongly correlated transition metal oxides due to the known
difficulty of defining the U parameter for the full composition range
and obtaining converged DFT+U energies  \cite{U-Eform,U-ramping}.

The computational cost of constructing and using NNs is reasonably low
and allows for modeling extended sets of multicomponent systems. The
DFT data generation stage requires an initial investment of about
10,000 CPU hours per unary, binary, or ternary set, with roughly one
CPU hour per small unit cell sized structure considered in this
study. For the described data and parameter set sizes, full NN
training takes close to 25 CPU hours for unaries and 600 CPU hours for
ternaries. Stratified NN training for a {\it single} ternary system
takes about half the total time ($3\times 25 + 3\times 50 + 100 = 325$
CPU hours). More importantly, construction of NN models for new sets
of multicomponent systems, e.g., all binary and ternary combinations
of 5 elements, should be faster by an order of magnitude. The speed-up
is significant even with the efficient parallelization of the training
process available in {\small MAISE}. As has been pointed out in
previous studies \cite{0953-8984-26-18-183001}, NNs are about factor
of 100 slower than traditional potentials but offer a considerable
speed-up compared to DFT calculations. In our implementation, a single
calculation of the energy, forces, and stresses for a 100-atom
structure takes a factor of $10^3$-$10^4$ less CPU time with NNs
compared to DFT.

It should be noted that while the NN 10-meV/atom accuracy might not be
sufficient to definitively resolve identified competing structures in
some systems, the coupling of the two bio-inspired algorithms can
dramatically reduce the pool of viable candidates that need to be
calculated at the DFT level. The structure search acceleration
strategy has been used previously in combination with pre-defined or
tuned-on-the-fly classical potentials \cite{jp1048088}. In the present
approach, a system-specific NN can be generated on demand. With the
available tests and monitors of the models' transferability, they may
be suitable for reuse in future large-scale simulations.

In summary, we have outlined and examined schemes for (i) automated
generation of data and (ii) stratified construction of multicomponent
NNs  which, to the best of our knowledge, have not been previously used
in the development of NN-based interatomic models. The employed data
generation protocol avoids certain biases present in the MD-based
approach and samples parts of the configurational space typically
accessed in ground state searches. The use of small-cell structures
enables creation of extended databases for fitting NNs with several
thousand adjustable weights. The stratified NN training procedure for
multicomponent systems is centered on reusing fitted parameters from
constituent NNs. Test results for a range of properties showed that
the accuracy of produced NNs is maintained for the considered Cu-Pd-Ag
system. The sequential fitting of parameters to corresponding
element-specific databases has the potential to accelerate the
construction of NN libraries and enable a direct comparison of
properties, such as formation energies, across chemical element
sets. It should be kept in mind that NNs, as any classical interatomic
models, are not universally transferable; therefore, unary NNs should
be carefully tuned and tested for targeted geometries, e.g., for bulk
or nanoparticle structure sets, before they can be expanded to treat
the corresponding multicomponent configurations. While NNs have
already been used for structure prediction \cite{Si-HP,Cu-Au-O-H-NP},
the methodology has yet to be widely adopted. Presented tests indicate
that NNs are robust enough to be used in evolutionary ground state
searches. With a moderate investment of computational resources to
build a NN for a specific alloy, one can explore the full composition
range using unconstrained optimizations which can in some cases take
over $10^5$ CPU hours for a single composition at the DFT level
\cite{ak23}. Application and expansion of the methodology to
accelerate structure prediction for a wider range of geometries
and chemical systems are the subject of our on-going work.

\section{Acknowledgments}
The authors gratefully acknowledge the NSF support (Award No. DMR-1410514).


\bibliography{ref}

\end{document}


\title{Supplementary Material to: Stratified construction of neural network-based interatomic models for multicomponent materials}

\author {Samad Hajinazar, Junping Shao, Aleksey N. Kolmogorov}

\affiliation{Department of Physics, Applied Physics and Astronomy, Binghamton University, State University of New York, PO Box 6000, Binghamton, New York 13902-6000, USA}

\date{\today}

\maketitle



\begin{widetext}
\begin{center}
 \begin{table*}[ht!]
 \hfill{}
\begin{tabular}{p{1cm}p{14cm}p{4cm}}

I    &  Fig. S1 Equation of state (EOS) results for NNs fitted without EOS data       \dotfill      & 1 \\
    & & \\

II    &  Fig. S2 NN errors for binary SQS and random structures in fcc and bcc lattices       \dotfill      & 2 \\
    & & \\

III  &  Fig. S3 Phonon dispersion tests for B2-CuAg       \dotfill      & 2 \\
    & & \\

IV & Fig. S4 Convergence of various properties during NN training                        \dotfill     & 3 \\ 
& & \\

V &  Table S1 Behler-Parrinello symmetry function parameters                      \dotfill      & 4 \\
    &  & \\

VI  &  Fig. S5 Distributions and correlations of weights in simplified binary NNs    \dotfill      & 5

    \end{tabular}
\hfill{}
\end{table*}
\end{center}
\end{widetext}

\begin{figure}[hb!]
\begin{center}
\includegraphics[width=170mm,angle=0]{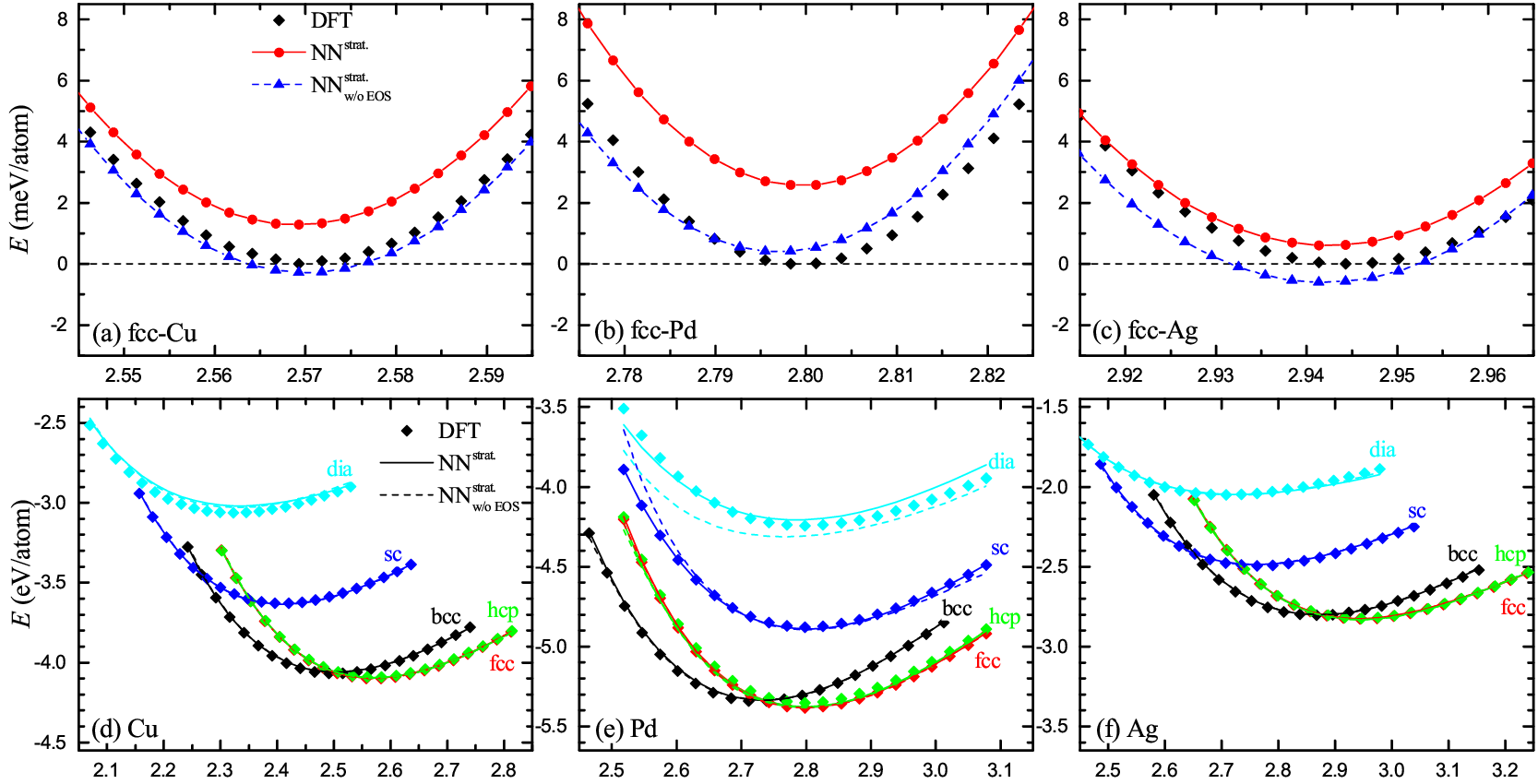}
\caption{Equation of states (EOS) for low-energy Cu, Pd, and Ag
  structures evaluated with NN$^{\text{strat.}}$ and
  NN$^{\text{strat.}}_{\text{w/o EOS}}$. The latter NN was trained
  without the EOS data.}
\vspace{0cm}
\end{center}
\end{figure}

\clearpage
\newpage

\begin{figure}[hb!]
\begin{center}
\includegraphics[width=150mm,angle=0]{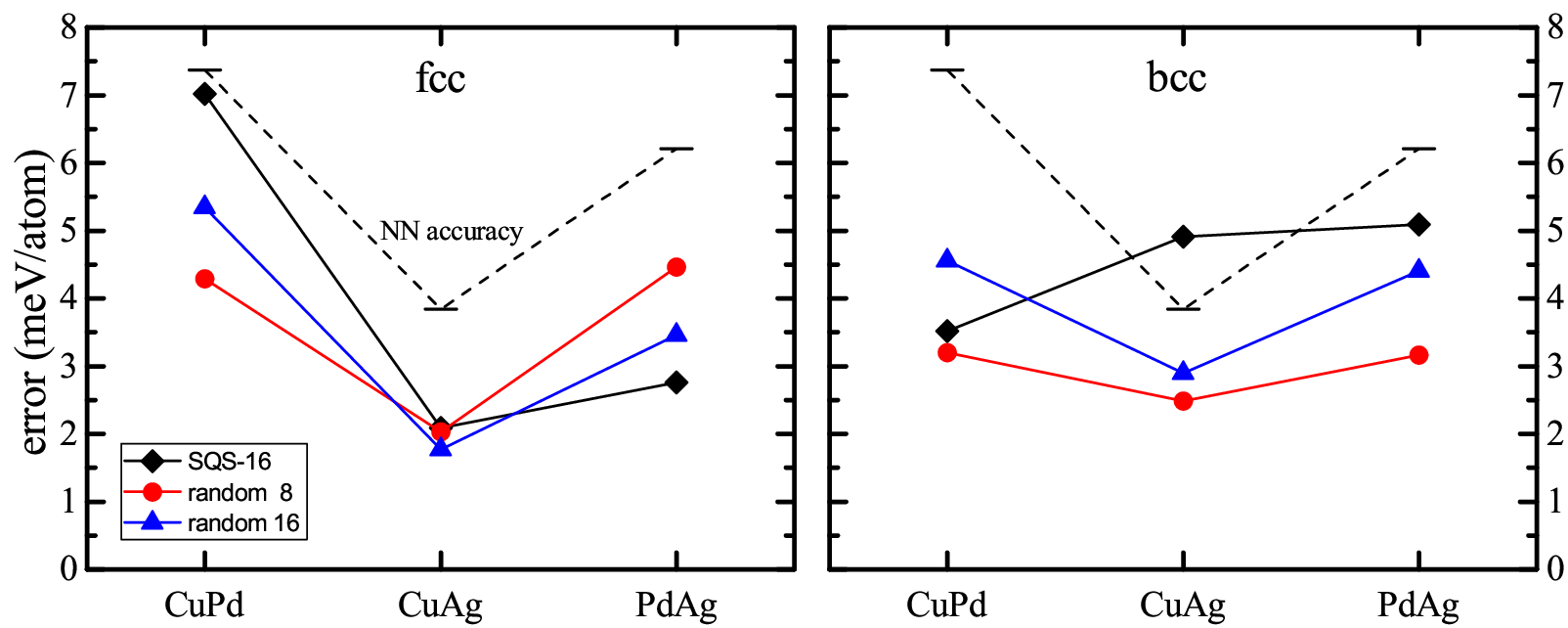}
\caption{NN total energy errors for solid solution alloys simulated
  with special quasirandom structures (SQS) and sets of random
  structures. All configurations were fully relaxed at the DFT
  level. The 16-atom SQS at the 25\%, 50\%, and 75\% compositions were
  taken from Ref. [80] (for the bcc lattice) and Ref. [81] (for the
  fcc lattice). Our bcc and fcc random structures were generated as
  differently decorated 8- and 16-atom supercells; after exclusion of
  identical configurations 13 8-atom and 60 16-atom distinct
  structures were considered across the full composition range for
  each lattice type. For comparison, we show the accuracy of the
  corresponding binary NNs as horizontal bars connected with dashed
  lines.}
\vspace{0cm}
\end{center}
\end{figure}

\begin{figure}[hb!]
\begin{center}
\includegraphics[width=120mm,angle=0]{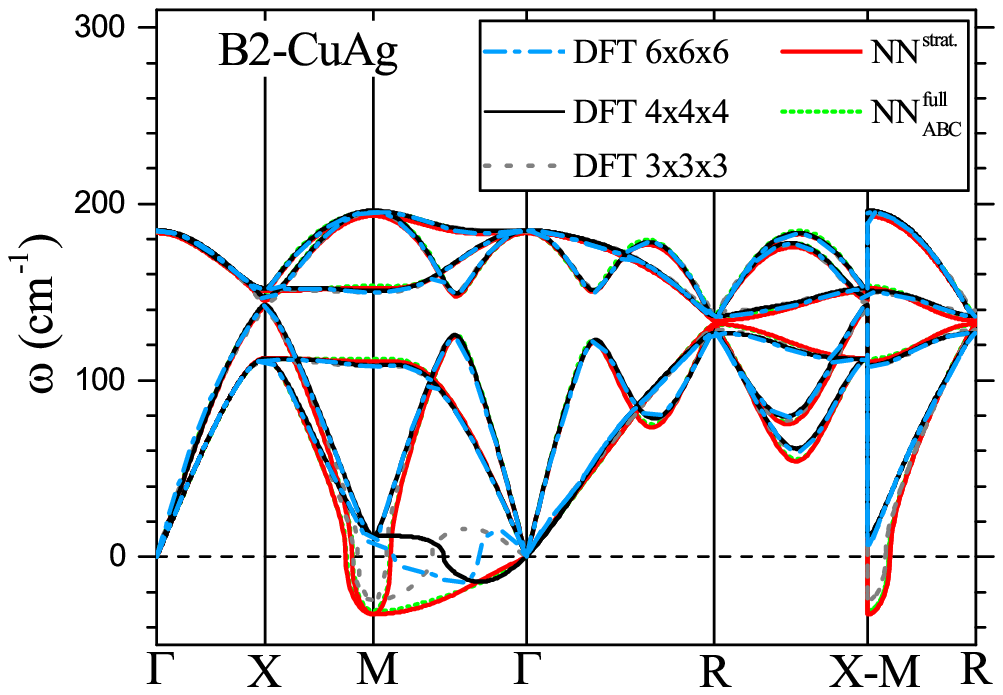}
\caption{Phonon dispersions for B2-CuAg calculated at the DFT and NN
  levels with the finite displacement method as described in Section
  VI.D. The plot illustrates the sensitivity of the DFT results for
  the soft mode frequencies to the size of the supercell. The
  discrepancies are seen for both the $M$-point frequency and the
  slope of the soft branch at $\Gamma$ along the $\Gamma-M$
  direction. The even $4\times4\times4$ (128 atoms, $4\times4\times4$
  k-mesh) and $6\times6\times6$ (432 atoms, $3\times3\times3$ k-mesh)
  expansions of the unit cell include the $M$-point explicitly giving
  $\omega_M$ values of 13 cm$^{-1}$ and 6 cm$^{-1}$, respectively. For
  comparison, a linear response calculation for the
  $\sqrt{2}\times\sqrt{2}\times1$ supercell ($12\times12\times18$ k-mesh)
  produced $\omega_M$ = 23 cm$^{-1}$. The considerably different DFT
  results for the soft branch could be caused in part by the
  application of the acoustic sum rule as discussed in our
  Ref. [84].}
\vspace{0cm}
\end{center}
\end{figure}

\clearpage
\newpage

\begin{figure}[ht!]
\begin{center}
\includegraphics[width=100mm,angle=0]{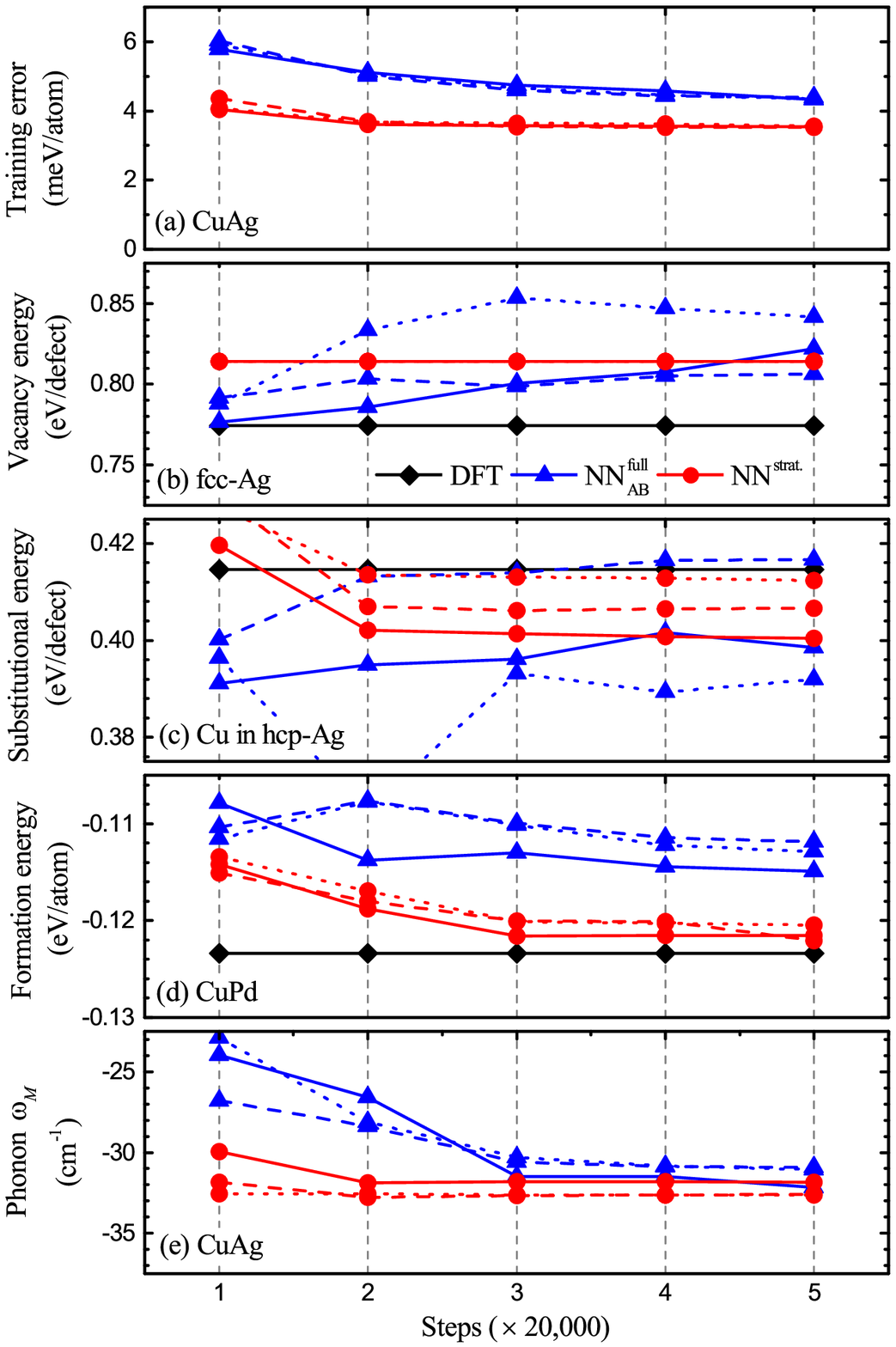}
\caption{Convergence test for NN performance as a function of number
  of training steps for full (blue up triangles) and stratified (red
  circles) NN results compared to DFT (black diamonds) data. Based on
  these tests, good convergence is achieved for all considered
  properties within 60,000 steps. (a) Training error for
  NN$_{\text{CuAg}}$. (b) Vacancy formation energy for Ag calculated
  using full and stratified NN$_{\text{CuAg}}$; the binary NNs trained
  in the stratified fashion inherit the same elemental parameters and
  give identical values. (c) Substitution energy of a Cu atom in Ag
  hcp structure. Stratified NN$_{\text{CuAg}}$ are evalutaing the
  substitution differently because binary parameters are adjusted to
  different arrangement of CuAg datasets (d) Formation energy of CuPd
  1:1 ratio ground state found in {\it{ab initio}} level evolutionary
  search vs. DFT formation energy (e) M point frequency of CuAg-B2
  structure reproduced with full and stratified NN$_{\text{CuAg}}$ }

\vspace{0cm}
\end{center}
\end{figure}

\begin{widetext}
\begin{center}
 \begin{table*}[ht!]
 \hfill{}
\begin{tabular}{cccccccccccccc}\hline\hline
&No.   &Type     & Rc    & Rs          & $\eta$      & $\zeta$ & $\lambda$ \\ \hline\hline

 &1  &pair     &6.0  &0.0000           &0.0010       &1.0000   &-1.0000 \\
 &2  &pair     &6.0  &0.0000           &0.0100       &1.0000   &-1.0000 \\
 &3  &pair     &6.0  &0.0000           &0.0200       &1.0000   &-1.0000 \\
 &4  &pair     &6.0  &0.0000           &0.0350       &1.0000   &-1.0000 \\
 &5  &pair     &6.0  &0.0000           &0.0600       &1.0000   &-1.0000 \\
 &6  &pair     &6.0  &0.0000           &0.1000       &1.0000   &-1.0000 \\
 &7  &pair     &6.0  &0.0000           &0.2000       &1.0000   &-1.0000 \\
 &8  &pair     &6.0  &0.0000           &0.4000       &1.0000   &-1.0000 \\ \hline
 &9  &triplet  &6.0  &0.0000           &0.0001       &1.0000   &-1.0000 \\
 &10 &triplet  &6.0  &0.0000           &0.0001       &1.0000   & 1.0000 \\
 &11 &triplet  &6.0  &0.0000           &0.0001       &2.0000   &-1.0000 \\
 &12 &triplet  &6.0  &0.0000           &0.0001       &2.0000   & 1.0000 \\
 &13 &triplet  &6.0  &0.0000           &0.0030       &1.0000   &-1.0000 \\
 &14 &triplet  &6.0  &0.0000           &0.0030       &1.0000   & 1.0000 \\
 &15 &triplet  &6.0  &0.0000           &0.0030       &2.0000   &-1.0000 \\
 &16 &triplet  &6.0  &0.0000           &0.0030       &2.0000   & 1.0000 \\
 &17 &triplet  &6.0  &0.0000           &0.0080       &1.0000   &-1.0000 \\
 &18 &triplet  &6.0  &0.0000           &0.0080       &1.0000   & 1.0000 \\
 &19 &triplet  &6.0  &0.0000           &0.0080       &2.0000   &-1.0000 \\
 &20 &triplet  &6.0  &0.0000           &0.0080       &2.0000   & 1.0000 \\
 &21 &triplet  &6.0  &0.0000           &0.0150       &1.0000   &-1.0000 \\
 &22 &triplet  &6.0  &0.0000           &0.0150       &1.0000   & 1.0000 \\
 &23 &triplet  &6.0  &0.0000           &0.0150       &2.0000   &-1.0000 \\
 &24 &triplet  &6.0  &0.0000           &0.0150       &2.0000   & 1.0000 \\
 &25 &triplet  &6.0  &0.0000           &0.0150       &4.0000   &-1.0000 \\
 &26 &triplet  &6.0  &0.0000           &0.0150       &4.0000   & 1.0000 \\
 &27 &triplet  &6.0  &0.0000           &0.0150       &16.000   &-1.0000 \\
 &28 &triplet  &6.0  &0.0000           &0.0150       &16.000   & 1.0000 \\
 &29 &triplet  &6.0  &0.0000           &0.0250       &1.0000   &-1.0000 \\
 &30 &triplet  &6.0  &0.0000           &0.0250       &1.0000   & 1.0000 \\ \hline
 &31 &triplet  &6.0  &0.0000           &0.0250       &2.0000   &-1.0000 \\
 &32 &triplet  &6.0  &0.0000           &0.0250       &2.0000   & 1.0000 \\
 &33 &triplet  &6.0  &0.0000           &0.0250       &4.0000   &-1.0000 \\
 &34 &triplet  &6.0  &0.0000           &0.0250       &4.0000   & 1.0000 \\
 &35 &triplet  &6.0  &0.0000           &0.0250       &16.000   &-1.0000 \\
 &36 &triplet  &6.0  &0.0000           &0.0250       &16.000   & 1.0000 \\
 &37 &triplet  &6.0  &0.0000           &0.0450       &1.0000   &-1.0000 \\
 &38 &triplet  &6.0  &0.0000           &0.0450       &1.0000   & 1.0000 \\
 &39 &triplet  &6.0  &0.0000           &0.0450       &2.0000   &-1.0000 \\
 &40 &triplet  &6.0  &0.0000           &0.0450       &2.0000   & 1.0000 \\
 &41 &triplet  &6.0  &0.0000           &0.0450       &4.0000   &-1.0000 \\
 &42 &triplet  &6.0  &0.0000           &0.0450       &4.0000   & 1.0000 \\
 &43 &triplet  &6.0  &0.0000           &0.0450       &16.000   &-1.0000 \\
 &44 &triplet  &6.0  &0.0000           &0.0450       &16.000   & 1.0000 \\
 &45 &triplet  &6.0  &0.0000           &0.0800       &1.0000   &-1.0000 \\
 &46 &triplet  &6.0  &0.0000           &0.0800       &1.0000   & 1.0000 \\
 &47 &triplet  &6.0  &0.0000           &0.0800       &2.0000   &-1.0000 \\
 &48 &triplet  &6.0  &0.0000           &0.0800       &2.0000   & 1.0000 \\
 &49 &triplet  &6.0  &0.0000           &0.0800       &4.0000   &-1.0000 \\
 &50 &triplet  &6.0  &0.0000           &0.0800       &4.0000   & 1.0000 \\
 &51 &triplet  &6.0  &0.0000           &0.0800       &16.000   &-1.0000 \\ \hline\hline

    \end{tabular}
\hfill{}
    \caption{Behler-Parrinello symmetry functions: set of functions
      and coefficients used in this paper in 8, 30 and 51 components
      sets}
\end{table*}
\end{center}
\end{widetext}

\clearpage
\newpage

\begin{figure}[ht!]
\begin{center}
\vspace{0.0 cm}
\includegraphics[width=87mm]{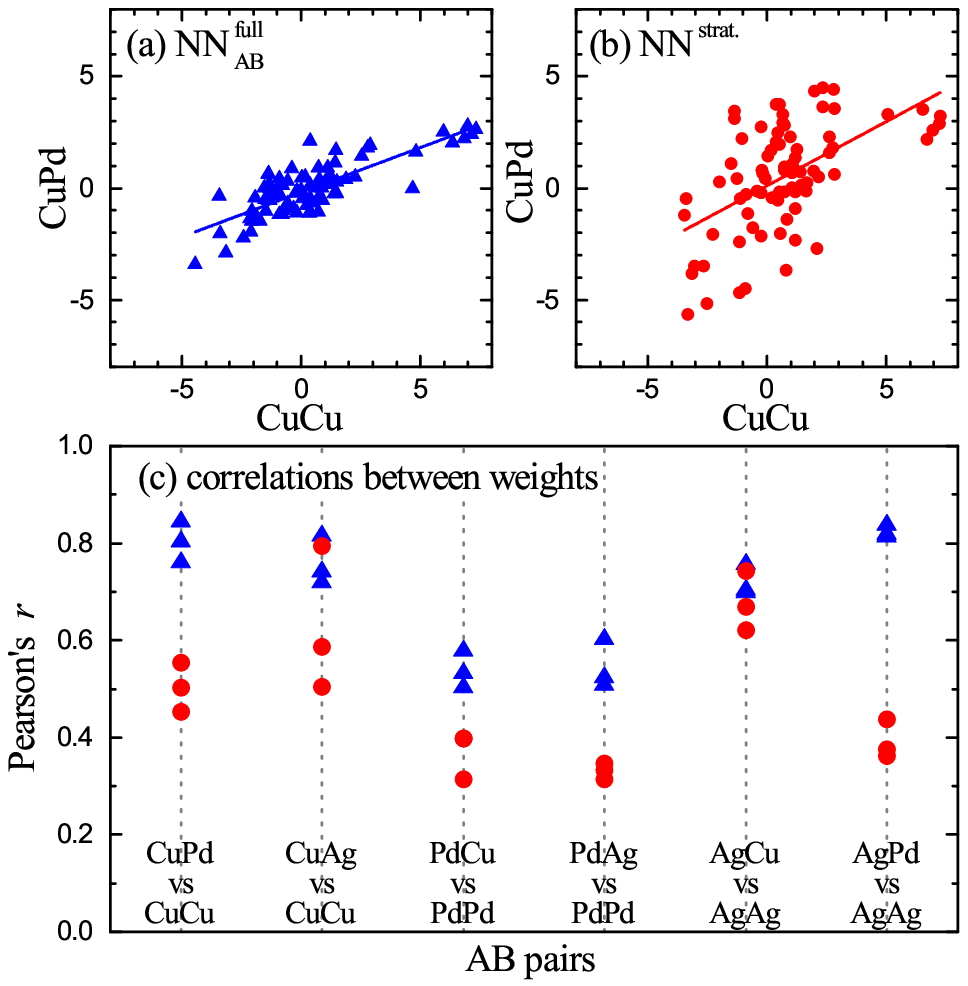}
\caption{Relative strength of weights for species AA (x-axis) and AB
  (y-axis) linking one of the 10 neurons in the first layer with the
  corresponding $G^1$ pair symmetry function in the input
  layer. 16-10-10-1 binary NNs with 8 pair symmetry functions per
  species were trained in the stratified fashion.}
\end{center}
\end{figure}